\documentclass[aps,prl,twocolumn,superscriptaddress,showpacs]{revtex4}

\usepackage{graphicx}

\begin{document}

% Use the \preprint command to place your local institutional report
% number in the upper righthand corner of the title page in preprint mode.
% Multiple \preprint commands are allowed.
% Use the 'preprintnumbers' class option to override journal defaults
% to display numbers if necessary
%\preprint{}

%Title of paper
\title{Laser-trapping of $^{225}$Ra and $^{226}$Ra with repumping by
room-temperature blackbody radiation}

% repeat the \author .. \affiliation  etc. as needed
% \email, \thanks, \homepage, \altaffiliation all apply to the current
% author. Explanatory text should go in the []'s, actual e-mail
% address or url should go in the {}'s for \email and \homepage.
% Please use the appropriate macro foreach each type of information

% \affiliation command applies to all authors since the last
% \affiliation command. The \affiliation command should follow the
% other information
% \affiliation can be followed by \email, \homepage, \thanks as well.

\author{J.~R.~Guest}
\affiliation{Physics Division, Argonne National
Laboratory, Argonne, Illinois 60439}

\author{N.~D.~Scielzo}
\affiliation{Physics Division, Argonne National
Laboratory, Argonne, Illinois 60439}
 
\author{I.~Ahmad}
\affiliation{Physics Division, Argonne National
Laboratory, Argonne, Illinois 60439}
 
\author{K.~Bailey} 
\affiliation{Physics Division, Argonne National
Laboratory, Argonne, Illinois 60439}
 
\author{J.~P. Greene}
\affiliation{Physics Division, Argonne National
Laboratory, Argonne, Illinois 60439}
 
\author{R.~J.~Holt}
\affiliation{Physics Division, Argonne National
Laboratory, Argonne, Illinois 60439}
 
\author{Z.-T.~Lu} \affiliation{Physics Division, Argonne National
Laboratory, Argonne, Illinois 60439} \affiliation{Department of
Physics and the Enrico Fermi Institute, University of Chicago,
Chicago, Illinois 60637}

\author{T.~P.~O'Connor}
\affiliation{Physics Division, Argonne National
Laboratory, Argonne, Illinois 60439}
 
\author{D.~H.~Potterveld}
\affiliation{Physics Division, Argonne
National Laboratory, Argonne, Illinois 60439}

%\email[]{Your e-mail address}
%\homepage[]{Your web page}
%\thanks{}
%\altaffiliation{}

%Collaboration name if desired (requires use of superscriptaddress
%option in \documentclass). \noaffiliation is required (may also be
%used with the \author command).
%\collaboration can be followed by \email, \homepage, \thanks as well.
%\collaboration{}
%\noaffiliation

\date{\today}

\begin{abstract}
We have demonstrated Zeeman slowing and capture of neutral $^{225}$Ra
and $^{226}$Ra atoms in a magneto-optical trap.  The intercombination
transition $^{1}S_{0}$$\rightarrow$$^{3}P_{1}$ is the only
quasi-cycling transition in radium and was used for laser-cooling and
trapping.  Repumping along the $^{3}D_{1}$$\rightarrow$$^{1}P_{1}$
transition extended the lifetime of the trap from milliseconds to
seconds.  Room-temperature blackbody radiation was demonstrated to
provide repumping from the metastable $^{3}P_{0}$ level.  We measured
the isotope shift and hyperfine splittings on the
$^{3}D_{1}$$\rightarrow$$^{1}P_{1}$ transition with the laser-cooled
atoms, and set a limit on the lifetime of the $^{3}D_{1}$ level based
on the measured blackbody repumping rate.  Laser-cooled and trapped
radium is an attractive system for studying fundamental symmetries.
\end{abstract}

% insert suggested PACS numbers in braces on next line
% insert suggested keywords - APS authors don't need to do this
%\keywords{}

%\maketitle must follow title, authors, abstract, \pacs, and \keywords
\maketitle

% body of paper here - Use proper section commands
% References should be done using the \cite, \ref, and \label commands
%\section{}
% Put \label in argument of \section for cross-referencing
%\section{\label{}}
%\subsection{}
%\subsubsection{}

Radium is an alkaline-earth element with no stable isotopes, and as a
result has eluded detailed spectroscopic studies.  Recent calculations
have suggested that radium, due to its nuclear and atomic properties,
is an excellent candidate for next generation fundamental symmetry
tests.  The observation of a permanent electric dipole moment (EDM) in
an atom, molecule, or the neutron would signal the long sought-after
violation of time-reversal symmetry ($T$-violation, or $CP$-violation
through the $CPT$ theorem) in a composite particle.  Currently, the
most stringent limits on $T$-violating interactions in the nucleus are
set by experiments that determined the atomic EDM of $^{199}$Hg to be
$<2.1 \times 10^{-28}$~$e$~cm~\cite{romPRL2001}.  Both
collective~\cite{auePRL1996} and
mean-field~\cite{engPRC2000} calculations have
predicted that radium isotopes which are characterized by nuclear
octupole deformation are two to three orders of magnitude more
sensitive to $T$-violating interactions in the nucleus than
$^{199}$Hg~\cite{dzuPRA2002,deJPRC2005}.  $^{225}$Ra is a particularly
promising candidate for an EDM search because it has a relatively long
half-life ($t_{1/2}=14.9$~days) and it has nuclear spin $I=1/2$ (like
$^{199}$Hg), which eliminates potential systematic electric quadrupole
shifts in EDM measurements.  In addition to the atomic
EDM, atomic parity violation effects are also enhanced by a few orders
of magnitude in radium due to the near degeneracy between the
parity-odd $^{3}P_{1}$ level and the parity-even $^{3}D_{2}$
level~\cite{flaPRA1999,dzuPRA2000}, which are believed to be only 5
cm$^{-1}$ apart~\cite{moore}.  Conventional atomic-beam or vapor-cell
approaches to these EDM or parity-violation measurements are precluded
because of the scarcity and low vapor pressure of radium,
respectively.  Instead, a measurement on atoms laser-cooled and
confined in a far-off resonant optical dipole trap or lattice
offers a promising path~\cite{romPRA1999}.

In this Letter, we report on the first realization of laser-cooling
and trapping of $^{225}$Ra and $^{226}$Ra ($t_{1/2}$=1600~yr, I=0).
This is the heaviest and only the second element with no stable
isotopes (after francium~\cite{simPRL1996}) laser-trapped to
date.  Radium is unique among alkaline-earth elements in that the
weak, intercombination transition $^{1}S_{0}$$\rightarrow$$^{3}P_{1}$
is the only quasi-cycling transition suitable for laser-cooling and
trapping.

\begin{figure}
  \includegraphics[scale=0.51]{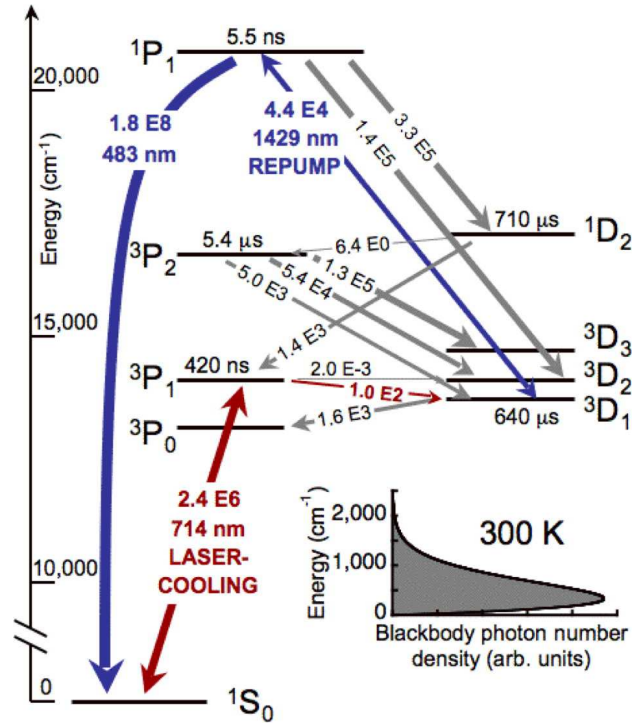}
  \caption{(Color online) Energy level diagram and decay rates for radium.
  Energies~\cite{moore} are indicated at the left, and calculated
  decay rates~\cite{dzuPRA2006} are shown ($^{3}P_{1}$
  lifetime comes from experiment~\cite{sciPRA2006}).  The 300~K
  blackbody photon number density is shown on the same energy
  scale in the inset.
  \label{radiag}}
\end{figure}

The energy level structure~\cite{moore} and decay
rates~\cite{dzuPRA2006} for the radium atom are shown in
Fig.~\ref{radiag}.  While the strong, singlet to singlet transition
$^{1}S_{0}$$\rightarrow$$^{1}P_{1}$ is exploited for efficient Zeeman
slowing and first-stage cooling in Mg, Ca, Sr, and Yb, the higher-$Z$
Ra atom is limited to $\approx$350 cycles on this transition before it
decays to long-lived $D$ levels.  In contrast, the narrow
intercombination $7s^{2}$~$^{1}S_{0}$$\rightarrow$~$7s7p$~$^{3}P_{1}$
transition is calculated to afford $\approx 2.4 \times 10^{4}$ cycles
before leaking to the $7s6d$~$^{3}D_{1}$
level~\cite{sciPRA2006,dzuPRA2006}.  By then repumping the atom to the
$7s7p$~$^{1}P_{1}$ level, which quickly decays back to the ground
$7s^{2}$~$^{1}S_{0}$ level, we extend the number of laser-cooling
transitions in the magneto-optical trap (MOT) to $\approx 3 \times
10^{7}$ and the laser-cooling time from tens of milliseconds to tens
of seconds.  As indicated by the spectrum inset in Fig.~\ref{radiag},
room temperature blackbody radiation can also redistribute population
between the $^{3}P_{1}$, $^{3}D_{2}$, $^{3}D_{1}$, and $^{3}P_{0}$
levels.

$^{226}$Ra is a long-lived isotope and $^{225}$Ra is an
$\alpha$-daughter of long-lived $^{229}$Th ($t_{1/2}=7340$~yr).
Typically, $\approx$0.7~$\mu$Ci of $^{226}$Ra ($\approx$0.7~$\mu$g)
and $\approx$1~mCi of $^{225}$Ra ($\approx$25~nano-g) are dissolved in
0.1 M nitric acid, pipetted onto an Al foil and placed in an oven with
$\approx$50~mg of Ba metal.  The Ba serves the dual purpose of
reducing the Ra(NO$_{3}$)$_{2}$ and passivating the oven surfaces.  We
typically run the oven at 600$^{\circ}$C - 750$^{\circ}$C and maintain
a vacuum in the trap region of $\approx 10^{-8}$~Torr.

For laser-cooling along the $^{1}S_{0}$$\rightarrow$$^{3}P_{1}$
transition ($\lambda = 714$~nm, $\Gamma = 2 \pi \times 380$~kHz,
$I_{sat}=140$~$\mu$W/cm$^{2}$), we excite along
$(J=0)$$\rightarrow$$(J'=1)$ for $^{226}$Ra and
$(F=1/2)$$\rightarrow$$(F'=3/2)$ for $^{225}$Ra.  We use a Ti:Sapphire
ring laser which is referenced to molecular iodine lines.  The repump
$^{3}D_{1}$$\rightarrow$$^{1}P_{1}$ transition is excited with an
external cavity diode laser at 1429~nm.  The laser is locked to a
Fabry-Perot cavity, which is stabilized by a He-Ne laser.  For
spectroscopy on the repump transition, the scanning repump laser is
monitored with a wavemeter and heterodyned with a second 1429~nm
external cavity diode laser which is locked to the Fabry-Perot.

The atoms emitted from the oven are first laser-cooled transversely in
two dimensions and then slowed in a 0.9~m Zeeman slower, which can
capture atoms with velocities up to 60~m/s ($10^{-3}$ of the
longitudinal velocity profile for $T\approx 700^{\circ}$C).  Because
our slow cycling transition and 3~cm laser beams yield a relatively
low MOT capture velocity of 10 m/s, we used an integrated geometry in
which the slower magnetic fields merge smoothly with the 1~G/cm MOT
magnetic field.  The transverse-cooling, slowing, and MOT beams are
typically detuned from the atomic transition by -3$\times$,
-12$\times$, and -6$\times \Gamma$ with intensities of 50$\times$,
30$\times$, and 20$\times I_{sat}$, respectively.

After slowing and collecting atoms in the trap for $\approx$0.5~s, the
transverse-cooling and slowing beams are shuttered and the MOT light
is frequency-shifted (typically to -3$\times \Gamma$) and attenuated
(to 2$\times I_{sat}$ total in all beams).  By chopping the
MOT light at 1~MHz (50\% duty cycle) and counting the fluorescence
photons while the light is off, we take advantage of the long
lifetime of the $^{3}P_{1}$ to eliminate scattered laser light and
gain sensitivity to single trapped radium atoms.  We can also detect 
the 483~nm photon from the $^{1}P_{1}$$\rightarrow$$^{1}S_{0}$ decay 
to monitor the repump rate.

\begin{figure}
  \includegraphics[scale=0.24]{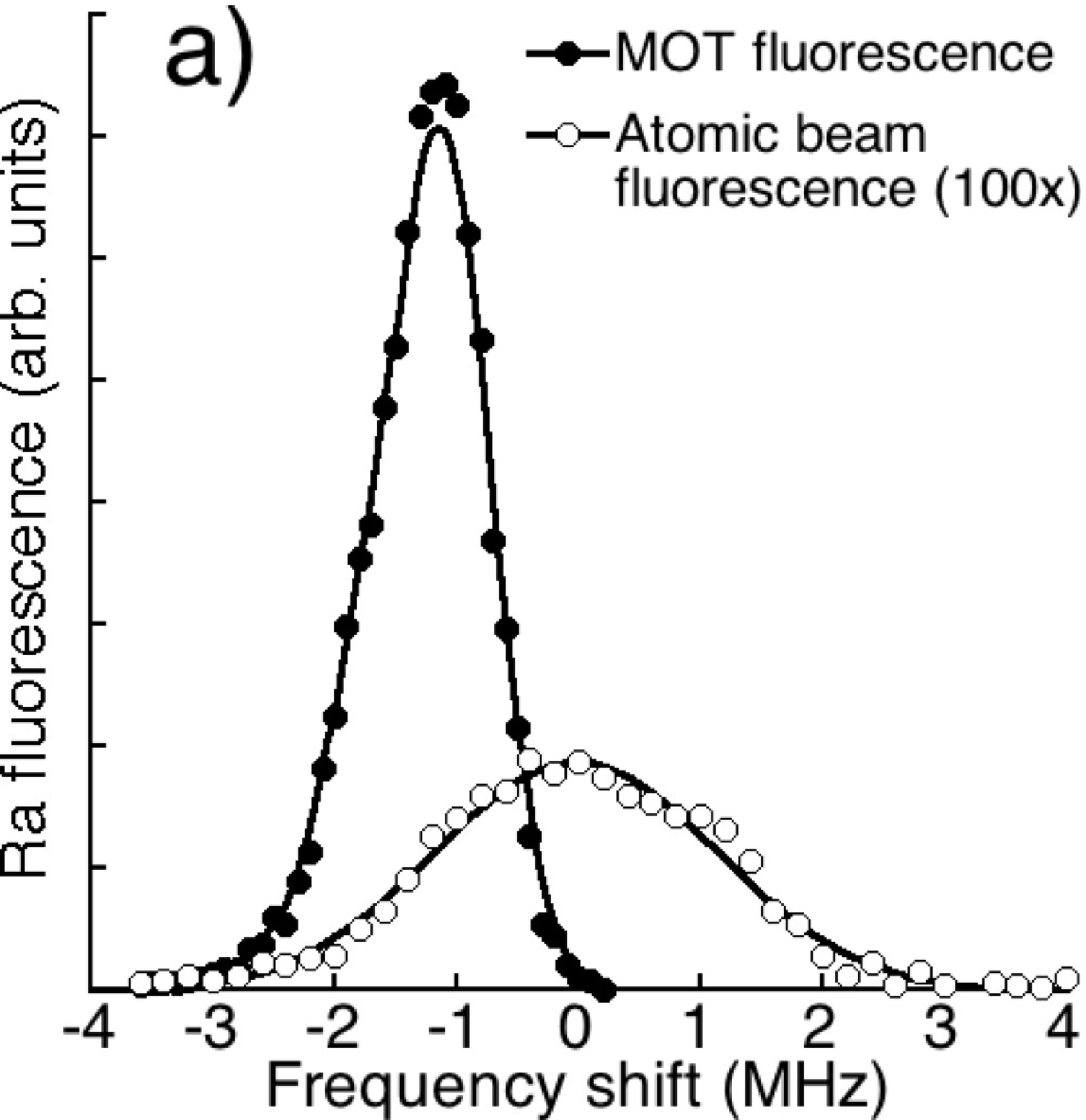}
  \includegraphics[scale=0.24]{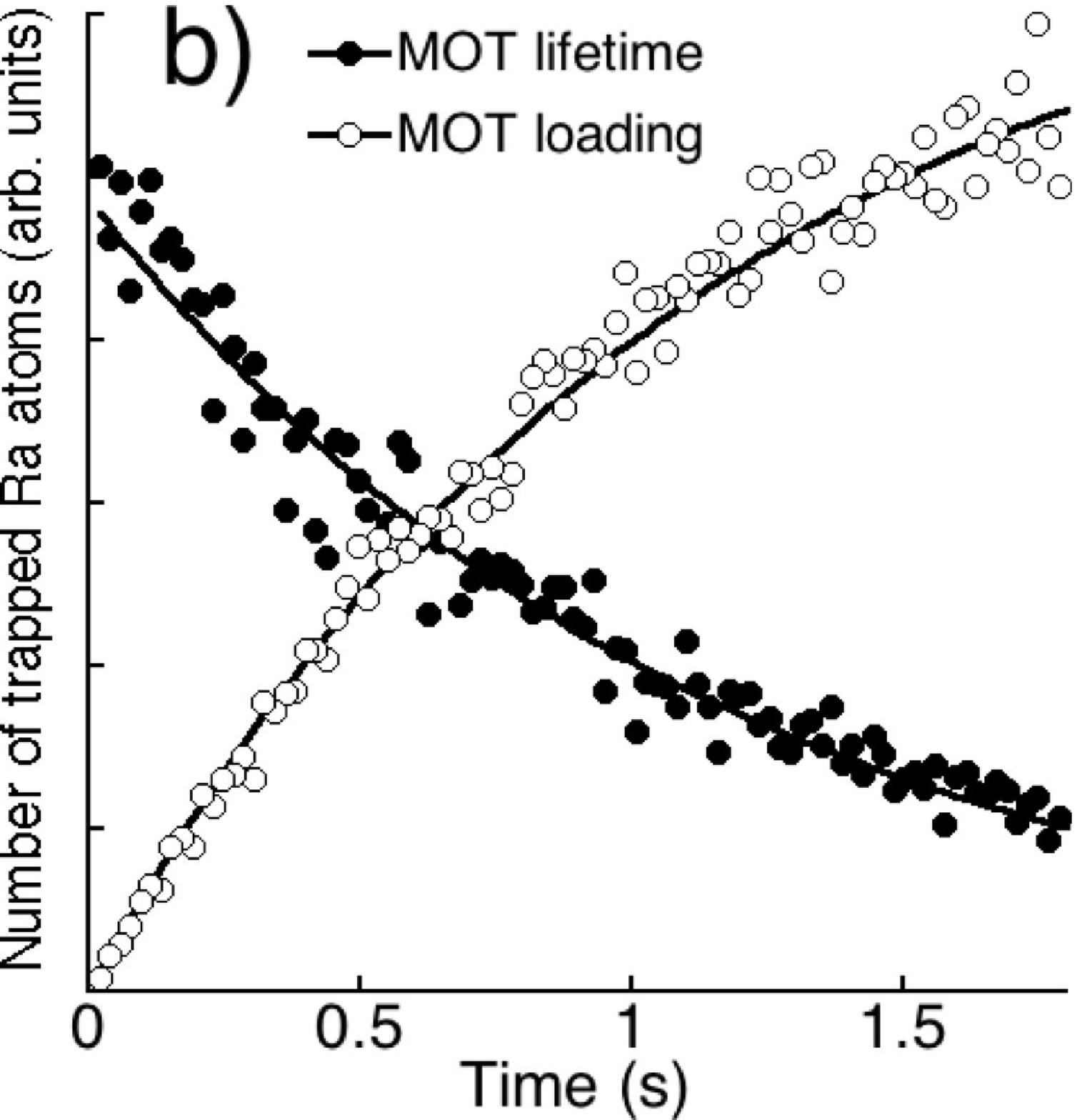}
  \caption{Laser-trapping of $^{226}$Ra atoms.  a) Spectra of the
  atoms in the atomic beam (multiplied by $100\times$, open circles)
  and laser-cooled in the MOT (filled circles).  b) The number of
  atoms detected in the MOT over the 1.8~s following the shuttering
  (unshuttering) of the slower beam at $t=0$ yields the lifetime
  (loading) curve which is shown by filled (open) circles.  Both
  curves are fit by exponentials and yield the identical lifetime
  $\tau = 1.1$~s for these data.
  \label{trapdata}}
\end{figure}

In Fig.~\ref{trapdata}a, we show the fluorescence from $^{226}$Ra
atoms in the trap chamber as a function of MOT light frequency for
both the untrapped atomic beam (multiplied by 100$\times$) and the
laser-cooled atoms captured in the MOT. During this probe phase, the
laser is still cooling as evidenced by the asymmetric line shape and
the red-shift of the MOT signal with respect to the atomic beam
signal.  Figure~\ref{trapdata}b shows exponential fits to the loading
and decay curves for the radium MOT, which yield a lifetime
$\tau=1.1$~s that is consistent with the time scale for collisions
with the background gas.  The loading efficiency of the MOT can be
defined as $\epsilon = L/F = N / (F \tau) $, where $L$ is the loading
rate, $F$ is the atomic flux from the oven, and $N$ is the number of
atoms observed in the trap.  The atomic flux $F$ of $^{225}$Ra is
obtained by exposing a deposition target to the atomic beam and
subsequently counting {\sl in situ} the number of 40~keV gamma-rays
emitted during the radioactive decay of the $^{225}$Ra nuclei using a
germanium detector; the $^{226}$Ra flux is estimated by scaling from
the $^{225}$Ra flux by the quantities introduced into the oven.
Typically, we observe fluxes of $3 \times 10^{7}$~s$^{-1}$ and
$10^{9}$~s$^{-1}$ and atom trap numbers of 20 and 700 for $^{225}$Ra
and $^{226}$Ra, respectively, yielding approximate efficiencies of
$\epsilon \approx 7 \times 10^{-7}$ for both isotopes.

\begin{figure}
  \includegraphics[scale=0.34]{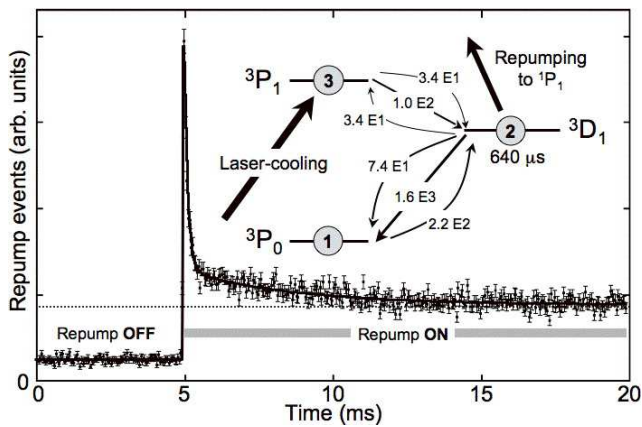}
  \caption{Repump events (483~nm photons) observed versus time as the
  repump laser is shuttered and unshuttered
  (laser-cooling is on continuously).  Experimental data (filled
  circles) and a double exponential fit (solid line for $t>5$~ms) are
  shown with a dotted line indicating the steady-state repumping rate.
  The inset shows the relevant levels and the calculated transition
  rates due to spontaneous emission (straight arrows) and thermal
  radiation (curved arrows).\label{BBdata}}
\end{figure}

Absolute frequencies of the laser-cooling
$^{1}S_{0}$$\rightarrow$$^{3}P_{1}$ transitions can be extracted based
on the known iodine line positions~\cite{iodineatlas,knoEPJD2004}.
The energies for $^{225}$Ra and $^{226}$Ra are
13999.269(1)~cm$^{-1}$~\cite{sciPRA2006} and 13999.357(1)~cm$^{-1}$,
respectively.  The value for $^{226}$Ra differs by 700~MHz from
Moore's value~\cite{moore} of 13999.38~cm$^{-1}$ which is based on the
1934 measurements by Rasmussen~\cite{rasZFP1934}.

As shown in the inset of Fig.~\ref{radiag}, thermal blackbody
radiation at 300~K should be expected to play an important role in
redistributing population between the $^{3}P_{1}$, $^{3}D_{2}$,
$^{3}D_{1}$, and $^{3}P_{0}$ levels.  Recognizing that our cold atoms
are sitting in a bath of room temperature thermal photons, we can
follow Einstein's simple treatment to relate the
spontaneous emission rates $A_{ji}$ to the thermal transition rates
$\tilde{B}_{ij}(T)=B_{ij}\rho(\omega_{ij},T)$, where
$\rho(\omega_{ij},T)$ is the blackbody photon energy density and
$B_{ij}$ is the Einstein $B$ coefficient.  For the levels in
the inset of Fig.~\ref{BBdata}, we have
\begin{equation}
    \tilde{B}_{ij}(T) =
    \frac{g_{j}}{g_{i}}\frac{A_{ji}}{e^{(E_{j}-E_{i})/k_{B}T}-1},
    \label{AB}
\end{equation}    
where $g_{i}$ is the degeneracy of the $i$th level, $E_{i}$ is the
energy of the $i$th level, and $k_{B}$ is the Boltzmann constant.
Taking the theoretical value for the decay along
$^{3}D_{1}$$\rightarrow$$^{3}P_{0}$ of $A_{21}=1.6 \times
10^{3}$~s$^{-1}$~\cite{dzuPRA2006} and the measured energy
difference of 638~cm$^{-1}$~\cite{moore}, we calculate a thermal
transition rate from $^{3}P_{0}$$\rightarrow$$^{3}D_{1}$ of
$\tilde{B}_{12}=2.2\times 10^{2}$~s$^{-1}$ for our chamber temperature
of $T = 296(3)$~K. Therefore, room temperature blackbody radiation can
drive population between these levels on millisecond time scales, as
indicated by the curved arrows in the inset of Fig.~\ref{BBdata}.

\begin{figure}
  \includegraphics[scale=0.35]{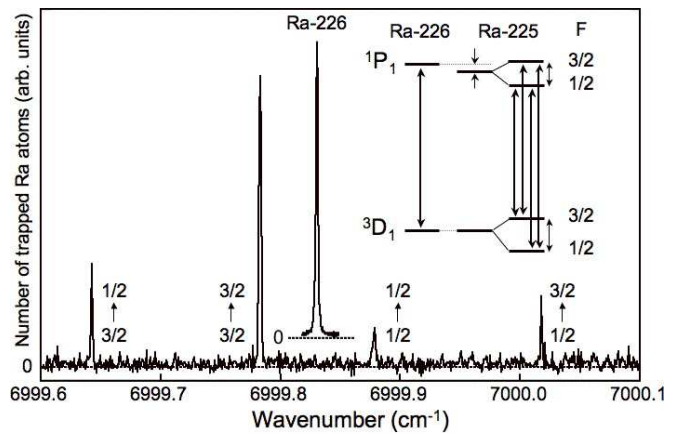}
  \caption{Number of $^{225}$Ra and $^{226}$Ra atoms in the trap, as
  measured by fluorescence at 714~nm, vs.\ the wavenumber of the
  1429~nm repump laser recorded by the wavemeter.  The $^{225}$Ra data
  is scaled by $100\times$ to compare to $^{226}$Ra, which is offset
  on the vertical scale for clarity.  The repump resonances are
  indicated in the inset level schematic.
  \label{rpspec}}
\end{figure}

This blackbody radiation mechanism can be seen directly in the data
shown in Fig.~\ref{BBdata}, where we show the observed rate of
repumping when shuttering and unshuttering the 1429~nm repump laser
(filled circles) while the MOT light (714~nm) is left on.  This
repumping rate is measured directly as a function of time by recording
the number of repump-induced 483~nm photons detected from the decay of
the $^{1}P_{1}$ level (as shown in Fig.~\ref{radiag}).  When the
repump laser is shuttered for the first 5~ms, population that leaks
during laser-cooling accumulates in the $^{3}D_{1}$ level and, after
radiative decay, in the metastable $^{3}P_{0}$ level.  When the repump
laser is reapplied to the cold atoms at $t=5$~ms, the atoms residing
in the $^{3}D_{1}$ level are quickly repumped and produce the short
burst of 483~nm photons seen in Fig.~\ref{BBdata}.  The slowly
decaying exponential tail following this initial pulse corresponds to
the emptying of the metastable $^{3}P_{0}$ level by thermal blackbody
radiation that reexcites the atoms to the $^{3}D_{1}$ level where they
can be repumped by the 1429~nm laser.  The signal then settles to a
steady-state repump rate.

These data are modeled effectively by a double exponential plus an
offset for $t>5$~ms, shown by the solid line in Fig.~\ref{BBdata}.
The time scale for the first exponential corresponds to the laser
repump rate and the time scale for the second exponential corresponds
to the blackbody repump rate.  This fit yields a blackbody repump rate
of $\tilde{B}_{12}=2.8\times 10^{2}$~s$^{-1}$, which, through
Eqn.~\ref{AB}, yields a lifetime for the $^{3}D_{1}$ level of $510 \pm
60$~$\mu$s for our chamber temperature.  We interpret this as a lower
limit on the lifetime because other loss mechanisms from the
$^{3}P_{0}$ level could contribute to the decay of this curve.

In practice, this blackbody mechanism eases the requirements on the
repumping laser fields because atoms which decay from the $^{3}D_{1}$
level to the metastable $^{3}P_{0}$ level will be recycled to the
$^{3}D_{1}$ level by the thermal radiation and be exposed to the
repumping laser again.  This is particularly significant for
$^{225}$Ra, which can be efficiently repumped by exciting from only
one of the hyperfine manifolds.  Assuming a uniform population of the
$^{3}P_{1} (F=3/2)$ $m_{F}$ states in the MOT, decay to the $^{3}D_{1}
(F=3/2)$ states is favored 5:1 over decay to the $^{3}D_{1} (F=1/2)$
states.  If we repump only the $^{3}D_{1} (F=3/2)$ states, atoms which
decay to the $^{3}D_{1} (F=1/2)$ states will further decay to the
metastable $^{3}P_{0}$ level.  But over the time scale of 4~ms, the
atoms will be recycled due to thermal radiation (as seen in
Fig.~\ref{BBdata} for $^{226}$Ra) to the $(F=3/2)$ states and
repumped.  This effect can also be seen in Fig.~\ref{rpspec}, which
shows the number of $^{225}$Ra and $^{226}$Ra atoms observed in the
trap as a function of the tuning of the repump laser; the four lines
for $^{225}$Ra correspond to $(F=1/2,3/2) \rightarrow (F'=1/2,3/2)$ as
indicated in the inset.  While most efficient along
$(F=3/2)\rightarrow (F'=3/2)$, repumping on any of the four transitions
leads to a significant increase (10$\times$-100$\times$) in the number
of trapped $^{225}$Ra atoms.

\begin{table}%[H] add [H] placement to break table across pages
\caption{Measured hyperfine structure constants $A$ in MHz for
$^{225}$Ra on the repump transition compared to a previous measurement
and theory (obtained by scaling calculations for $^{213}$Ra
\cite{dzuPRA2000} and $^{223}$Ra~\cite{biePRA2005} by the ratio of
nuclear magnetic moments and the inverse ratio of the nuclear
spins~\cite{arnPRL1987,ahmPhL1983}).\label{spectable}}
\begin{ruledtabular}
\begin{tabular}{lllcc}
% Lines of table here ending with \\
     & \multicolumn{2}{c}{Experiment} & \multicolumn{2}{c}{Theory} \\
    Level & This Letter & ISOLDE~\cite{ahmPhL1983} &
    \cite{dzuPRA2000} & \cite{biePRA2005}
    \\
    \hline
    $^{3}D_{1}$  &  4687.7 (1.5)  &   & 4915 & 4981 \\
    $^{1}P_{1}$  &  2797.3 (1.5)  &  2796.5 (2.5) & 1972 & 2688 \\
\end{tabular}
\end{ruledtabular}
\end{table}

By comparing the center frequencies of the $^{225}$Ra and $^{226}$Ra
lines, we have measured the isotope shift on the
$^{3}D_{1}$$\rightarrow$$^{1}P_{1}$ transition to be $\nu^{226} -
\nu^{225} = 540.2(2.0)$~MHz.  We have also extracted the hyperfine
structure constants $A$ for $^{225}$Ra, where the hyperfine energy
splitting $E(F=3/2)-E(F=1/2) = 3/2$~$h A$.  These values are presented
in Table~\ref{spectable} and compared with
theory~\cite{dzuPRA2000,biePRA2005}.  Our measurement agrees well with
the previous measurement of $A$ for the $^{1}P_{1}$ state at
ISOLDE~\cite{ahmPhL1983}.

Figure~\ref{rpspec} also shows that the
$^{3}D_{1}$$\rightarrow$$^{1}P_{1}$ repump transition energy in
$^{226}$Ra is measured to be 6999.84(2)~cm$^{-1}$.  This confirms
Russell's 1934 adjustment~\cite{rusPR1934} to the original arc
spectrum published by Rasmussen~\cite{rasZFP1934}, which shifted the
$D$-level and $F$-level energies by +628~cm$^{-1}$.  This measurement
resolves the conflict between calculations which were consistent with
the correction~\cite{eliPRA1996,dzuPRA2006} and theoretical concerns
that these levels could be much lower in energy~\cite{bieJPB2004},
which would have presented much stronger leak channels from the
$^{3}P_{1}$ level and therefore made laser-cooling significantly more
difficult.  It also suggests that the apparent near-degeneracy of the
$^{3}P_{1}$ and $^{3}D_{2}$ levels~\cite{moore} which would enhance
atomic parity violation effects in radium should be
correct~\cite{flaPRA1999,dzuPRA2000}.

The successful laser trapping of $^{225}$Ra and $^{226}$Ra has opened
the possibility of studying fundamental symmetries with cold, trapped
radium atoms.  Furthermore, to our knowledge, this is the first
demonstration that blackbody radiation can serve as an effective
repump source.  This mechanism may find more uses in laser trapping of
atoms with complex structure~\cite{mccPRL2006}.  In an EDM search, we
estimate that a statistical sensitivity of $10^{-26}$~$e$~cm can be
achieved using a 10~mCi $^{225}$Ra source with the demonstrated
trapping efficiency.  Due to radium's enhanced sensitivity to nuclear
$T$-violating interactions, a search with this level of sensitivity
would be competitive with the best current limits set in the nuclear
sector~\cite{romPRL2001}.

\begin{acknowledgments}
We would like to thank M.\ Williams and D.\ Bowers for technical
support and E.\ Schulte for early work on
the $^{225}$Ra oven.  We would also like to thank V.\ Flambaum, V.\
Dzuba, H.\ Gould, P.\ Mueller, K.\ Wendt and R.\ Santra for helpful
discussions.  This work was supported by the U.S.\ Department of
Energy, Office of Nuclear Physics, under Contract No.\
DE-AC02-06CH11357.
\end{acknowledgments}

% Create the reference section using BibTeX:

\end{document}